# Integrated and Steerable Vortex Lasers using Bound States in Continuum

B. Bahari, F. Vallini, T. Lepetit, R. Tellez-Limon, J. H. Park, A. Kodigala, Y. Fainman, and B. Kanté

*Department of Electrical and Computer Engineering, University of California San Diego, La Jolla, CA 92093-0407, USA*

bkante@ucsd.edu

**Abstract**-Orbital angular momentum is a fundamental degree of freedom of light that manifests itself even at the single photon level. The coherent generation and beaming of structured light usually requires bulky and slow components. Using wave singularities known as bound states in continuum, we report an integrated device that simultaneously generates and beams powerful coherent beams carrying orbital angular momentum. The device brings unprecedented opportunities in the manipulation of micro-particles and micro-organisms, and, will also find applications in areas such as biological sensing, microscopy, astronomy, and, high-capacity communications.

Vortices are recognized to be ubiquitous features around wave singularities, forming threads of silence in acoustic, turbulences in superfluids, and tidal waves in oceans[1,2]. Light carrying Orbital Angular Momentum (OAM) has been intensively investigated due to interest in numerous fields including metrology, microscopy, quantum entanglement, and particle trapping[3-6]. The theoretically unlimited orthogonal basis of OAM spatial modes makes it a candidate to overcome the data transmission capacity crunch[7]. In 1992, Allen *et al.* demonstrated that beams with non-zero OAM have a "doughnut"-shaped intensity profile and carry a discrete OAM of $l\hbar$ per photon that results from the $e^{il\Phi}$ dependence where $\Phi$ is the azimuthal angle, $\hbar$ the reduced Planck constant, and the integer *l* is called the topological charge[8]. On one hand, OAM in light is generally generated using non-integrated and unwieldy optical components that shape the phase front of a Gaussian laser beam. Generation methods include using angular grating[9], spiral phase plates[10], chiral dielectric meta-holograms[11], spatial light modulators[12], metasurfaces[13-16]. Recently a micro-ring laser operating at a parity-time-symmetry based Exceptional Point (EP) was demonstrated[17]. However, the strict requirement of EP operation severely limits the pump power to the single value ensuring the loss/gain balance, which in turn limits the output power of the laser. Many applications, including particle trapping and light detection and ranging (LIDAR), have minimum power requirements and would benefit from a scalable and integrable vortex beam source. On the other hand, beam steering, i.e., the ability to steer energy in controllable directions, has long been pursued and is usually implemented by optical phased arrays[18-20] using technologies such as liquid crystals[21], acousto-optics[22], and electro-optics[23,24]. Although Micro-Electro-Mechanical Systems (MEMS) have been recognized as a practical solution[25], state-of-the-art optical beam steering uses mechanisms that control the phase/intensity profile of individual nanoantennas for beam forming[18]. These mechanisms also suffer from limitations such as complexity, poor agility, and large size of non-integrated components. Furthermore, steering is usually limited to regular Gaussian beams. Here, we report an integrated photonic structure that is capable of simultaneously generating and beam coherent vortex beams in multiple and arbitrary directions controlled by the symmetry of the photonic structure. The vortex beam originates from a singularity where destructive interference occurs between radiative channels of the system[26-31]. The simultaneous generation and beaming of coherent vortex beams from a single, scalable, and integrated source constitutes a major step forward, as it enables the full and remote optical manipulation of micro-particles and micro-organisms not only by allowing their remote trapping and rotation but also by enabling their translation.

The schematic of the device is shown in Fig. 1a. It is made of structured InGaAsP multiple quantum wells epitaxially grown on an InP substrate, and tailored to emit around the telecommunication wavelength. The structure is fabricated by electron beam lithography followed by dry etching to form holes and thus constitutes a Photonic Crystal (PhC). The PhC is subsequently bonded on a flat glass substrate coated with a thin layer of PolyMethyl MethAcrylate (PMMA). During the bonding process, PMMA infiltrates holes of the PhC. Finally, the InP substrate is removed by wet etching using hydrochloric acid (see Supplementary Information). Figure 1b shows a top scanning electron micrograph view of a device, illustrating the uniformity of the holes in the structure.

To understand the physics of the device, modes of the PhC are calculated using finite element method and experimentally measured by pumping the device with a high-energy laser and collecting the photoluminescence at different angles around the wavelength of 1.6 µm (see Supplementary Information). Figure 1c presents the three modes within the gain bandwidth of InGaAsP, and, a very good agreement is observed between the theoretical and the experimental band diagrams.

Mode 1 is a singly degenerate mode while modes 2 and 3 are doubly degenerate modes. By changing the radius of holes uniformly over the whole structure, all three modes shift spectrally. The quality factor of mode 1 is theoretically infinite at the center of the Brillouin zone (Γ point) and independent of the radius of holes as mode 1 is symmetry protected (Γ-locked). This type of mode (usually referred to as band-edge mode) has been extensively investigated in uniform PhC lasers that necessarily emit normal to the surface[32]. Modes 2 and 3 can only have infinite quality factors if all radiation channels destructively interfere to form singular states, at reciprocal space points, known as bound states in continuum[33]. These singular states are robust vortex centers carrying non-zero topological charges[34-37], and a continuous change of parameters of the systems continuously tunes the destructive interference condition away from the Γ point, resulting in beaming of mode 2 along ΓM and mode 3 along ΓX (see Fig. 2a). Perturbing such modes, for example by varying the radius of holes of an array, is thus a method to beam laser beams without using additional components, such as phase arrays[13,18] or breaking the symmetry of the system to generate an artificial band-edge mode[38]. In our system, all holes have the same size in an array, and, the required phase shift to beam the light is naturally provided by the phase-offset between Floquet-Bloch harmonics of the periodic structure. It is worth noting that the broken $\sigma_z$ symmetry and the finite size of the realized samples limit the quality factor of modes 2 and 3 to very large but not infinite values, forming quasi-bound states in continuum (see Supplementary Information)[30,34,39].

Several devices with a range of radius of holes were fabricated to observe beaming. The measurements are performed using a micro-photoluminescence setup in which the reciprocal space is obtained by Fourier transforming the image plane (see Supplementary Information). The devices are optically pumped with a pulsed laser ($\lambda$ = 1064 nm, T = 12 ns pulse at a repetition rate f = 275 kHz). The evolution of the output power as a function of the pump power (light-light curve) for a representative sample confirms the threshold behavior and a clear transition from spontaneous emission to lasing (Fig. 2b). Figure 2c-f present reciprocal space images of the emission of four samples with decreasing radii. The center of the image represents the center of the Brillouin zone i.e., $k_x = k_y = 0$ ($\Gamma$ point). For R = 250 nm, the PhC supports only the $\Gamma$-locked mode. Figure 2c confirms normal emission from this sample. As the radius is decreased below R = 250 nm, the beaming of four lasing beams along $\Gamma$X is observed. They correspond mode 3, which has a large quality factor along $\Gamma$X, and the four beams stem from the four-fold symmetry of the crystal (invariance under 90° rotation). The number of beams can be controlled by the symmetry of the crystal as well as by boundary conditions (see Supplementary Information). The beaming angle further increases by decreasing the radius as seen in Fig. 2d-f. The emission angles are extracted from the reciprocal space images and the numerical aperture of the collecting objective. Blue dots in Fig. 2a represent the experimental emission angles, and, as can be seen, lasing occurs in directions of predicted high quality factors. The operating mode is selected by its spatial overlap with the optical gain and switches from mode 1 to mode 3 for R ~ 250 nm (see Supplementary Information).

Singular states resulting from full destructive interference carry a quantized topological charge (order of singularity) that can be controlled by the topology of the structure and mathematically refer to modes with singular far-field phase whose order can be identified by the far-field polarization twisting around the singular point in reciprocal space[34,35] or real space[40]. Figures 3a-f represent the experimental (A-C) and theoretical (D-F) far-fields of devices emitting at different angles. For R = 180 nm and R = 225 nm, mode 3 ($\Gamma$X-beaming), emits at an angle with a tilted doughnut-shaped pattern. For R = 250 nm, only mode 1 ($\Gamma$-locked), emits with a doughnut-shaped far-field.

To experimentally demonstrate the topology of these emissions, we perform self-interference of the lasing beams (see Supplementary Information). The interference patterns comprise two inverted fork-shaped patterns and thus demonstrate a topological charge of +1 and -1 for modes 3 and 1, respectively (Fig. 3g-i).

Singularities are always created in pairs (due to charge conservation)[35]. This is evident in Fig. 2a where a high quality factor mode that beams along $\Gamma$M is predicted. However, this mode was not observed so far as it experiences less effective gain than the $\Gamma$X-beaming mode (see Supplementary Information). Since the existence of this mode is guaranteed by topology, it should be possible to observe lasing and beaming along $\Gamma$M if the pump power is increased.

Figures 4a-f present the evolution of the reciprocal space image as a function of the pump power for samples supporting all three singular modes. For R = 190 nm, Fig. 4a clearly shows lasing along

ΓX. As the pump power is increased, four additional spots appear along ΓM (Fig. 4b-c). Similarly, for *R* = 225 nm, additional spots appear (at a smaller angle) as the pump power is increased (Fig. 4D-F). These results demonstrate beaming along ΓM. The modes can be further identified by their wavelength scaling. In Fig. 4g, the continuous lines represent the theoretical prediction of the wavelength of each mode with large quality factor as a function of the radius of holes. Dots represent measured wavelengths of emissions from the lasers and a good agreement with theory is observed.

We have demonstrated a topological light source that generates and beams coherent vortex beams from a fully integrated laser. The device, made of an array of holes in InGaAsP multiple quantum wells, operates at singular points where destructive interference occurs between radiative channels of the photonic structure. These points continuously move along lines in reciprocal space as a function of the radius of holes. The integrated system emits light with orbital angular momentum that can be beamed in arbitrary directions controlled by the symmetry of the crystal without requiring mechanical components or breaking the in-plane symmetry of the system for beam forming. Dynamical beam steering can be implemented by bonding the structure on electro-optic, acousto-optic, or phase-changed materials that will change the effective index of the slab and result in beam steering using a single control instead of $N^2$ controls in N-by-N arrays. The proposed scalable and steerable vortex beam source open unprecedented opportunities in micro-particles and micro-organisms manipulation as it enables not only their remote trapping and rotation from powerful emitted beams but also their translation. The device will find applications in multiple areas such as biological sensing, microscopy, astronomy, and high-capacity communications.

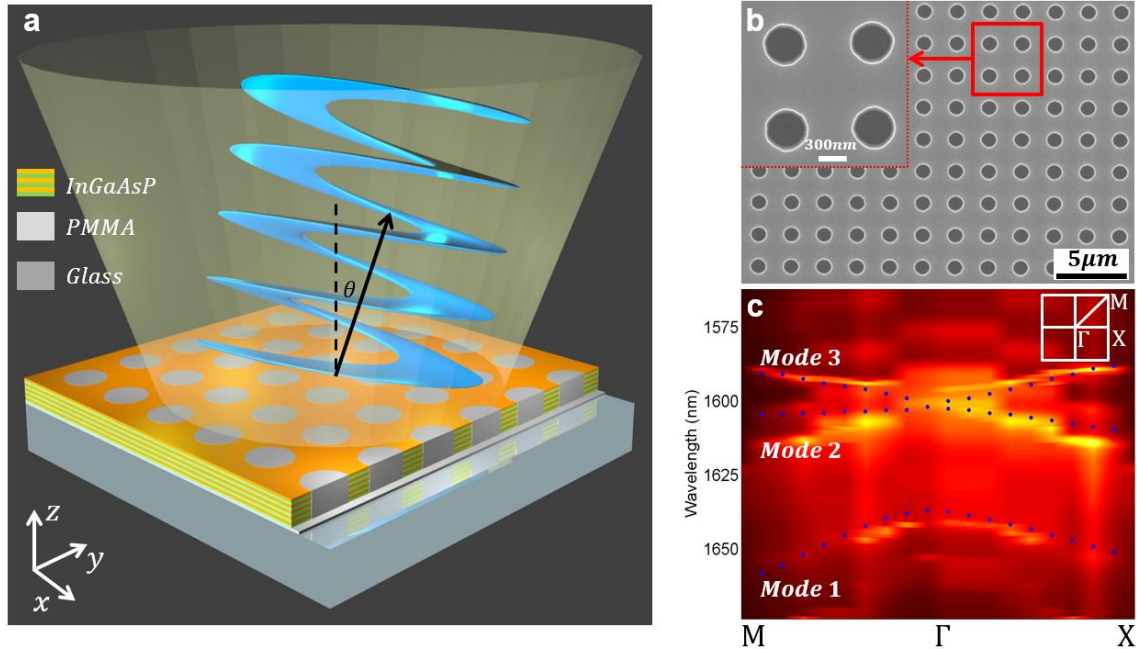

**Figure 1 | Steerable coherent vortex beam source.** (**a**) Schematic of the light source that is optically pumped (yellow beam) and generates a vortex beam at a controllable angle (blue beam). The structure is a square lattice photonic crystal (PhC) made of holes etched in InGaAsP multiple quantum wells bonded on a glass substrate coated with a thin layer of polymethyl methacrylate (PMMA). The holes are defined by electron beam lithography followed by wet etching. During the bonding process, PMMA infiltrates holes of the PhC. The InP substrate, on which InGaAsP is epitaxially grown, is subsequently removed by wet etching (see Supplementary Information). (**b**) Top view scanning electron micrograph of a fabricated structure with a quantum well thickness of 300 nm, a period of 950 nm, and holes radius of 200 nm. (**c**) Experimental (color plot) and theoretical (blue dots) band diagrams measured/calculated along the ΓX and the ΓM directions. The experimental band diagram is measured by pumping the structure with a high-energy laser and collecting the photoluminescence at different angles (see Supplementary Information). A good agreement is observed between the theoretical band diagram (using finite element method) and experiments.

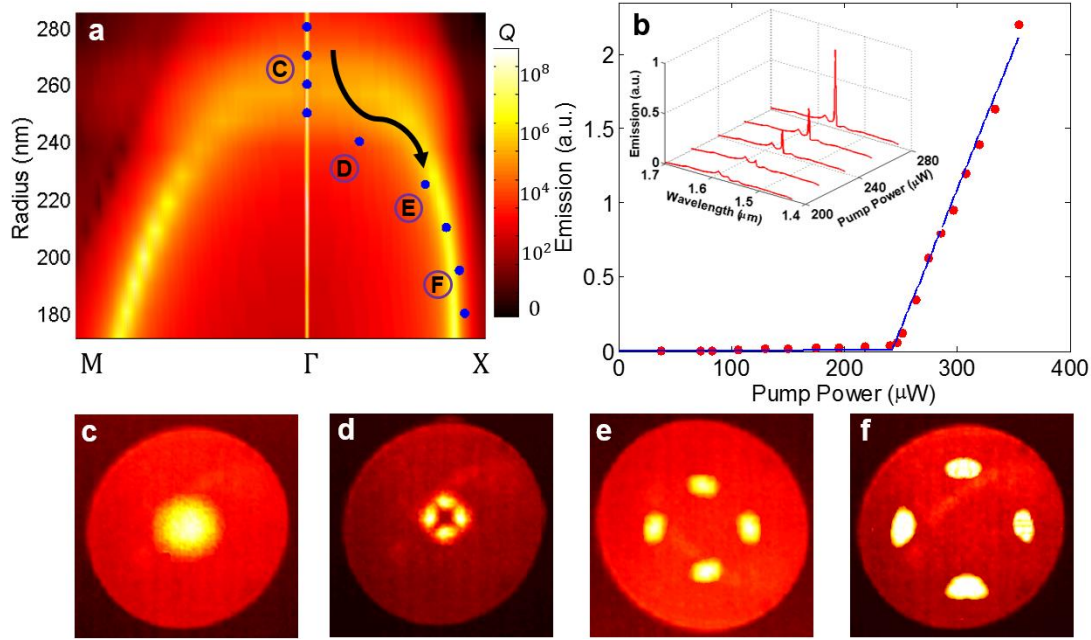

**Figure 2 | Demonstration of beaming in crystal's high-symmetry directions.** (**a**) Quality factor (Q) of modes 1-3 as a function of the radius of holes in reciprocal space (k-space). The quality factors of modes 1, 2 and 3 are singular in k-space, at a point for mode 1 (Γ-locked), and along lines for mode 2 (beaming along ΓM) and mode 3 (beaming along ΓX). Modes 2 and 3 can thus continuously beam with the radius of holes. The quality factor of mode 1 is singular for all radii of holes because mode 1 is symmetry-protected while mode 2 and 3 become singular for holes radii smaller than $R \sim 250$ nm. Blue dots correspond to experimental measurements of the beaming angle that have been extracted from k-space imaging (see Supplementary Information). The standard error in angle, from at least three measurements, is smaller than 0.5°. (**b**) Output power as a function of the average pump power (light-light curve) around the lasing wavelength for $R = 280$ nm. The red dots are experimental measurements of the output power for different pump powers. The blue solid lines are linear fits to the data in spontaneous and stimulated emission regimes and clearly show a threshold behavior, i.e. lasing. The inset shows the emission power evolution as a function of the pump power in a broader wavelength range. A similar light-light curve is measured for all other lasers (not shown). (**c-f**) Experimental k-space images of the emission from lasers with different radii. The distance of the bright spots to the center represents the in-plane wave-vector or equivalently the angle, and it increases as the radius of holes is decreased from $R = 250$ nm (c), to $R = 240$ nm (d), $R = 225$ nm (e), and $R = 190$ nm (f), clearly demonstrating beaming of the lasing beam. The maximum observable angle in k-space imaging is given by the numerical aperture (NA) of the objective lens (NA = 0.4). The four-fold symmetry of the structure implies that any singularity is invariant under 90° rotation and this is confirmed by the four observed bright spots creating a 2-by-2 array of optical vortices (see Fig. 3).

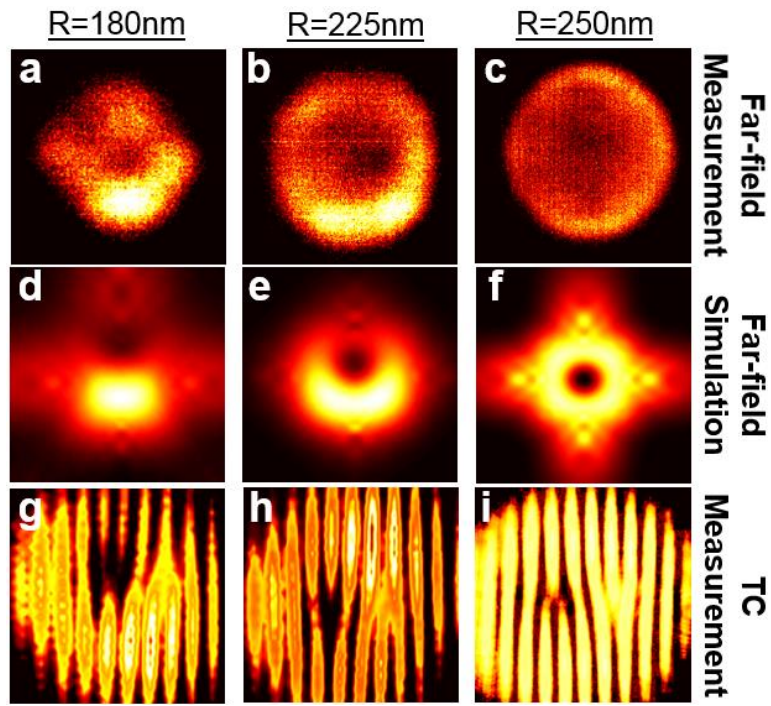

**Figure 3 | Real space image of the vortex beams and topological charge measurement.** Far-field radiation pattern measurements (**a-c**) and corresponding simulations (**d-f**) of the vortex lasers for three different radii, *R* = 180 nm (a, d), *R* = 225 nm (b, e), and *R* = 250 nm (c, f). By reducing the radius of holes, the far-field beam, which is doughnut-shaped, starts to beam. (**g-i**) Self-interference patterns of the emission beams for three samples with different radii. The two inverted fork-shape patterns demonstrate a topological charge (TC) of +1 for the mode beaming along ΓX (mode 3) and -1 for the Γ-locked mode (mode 1).

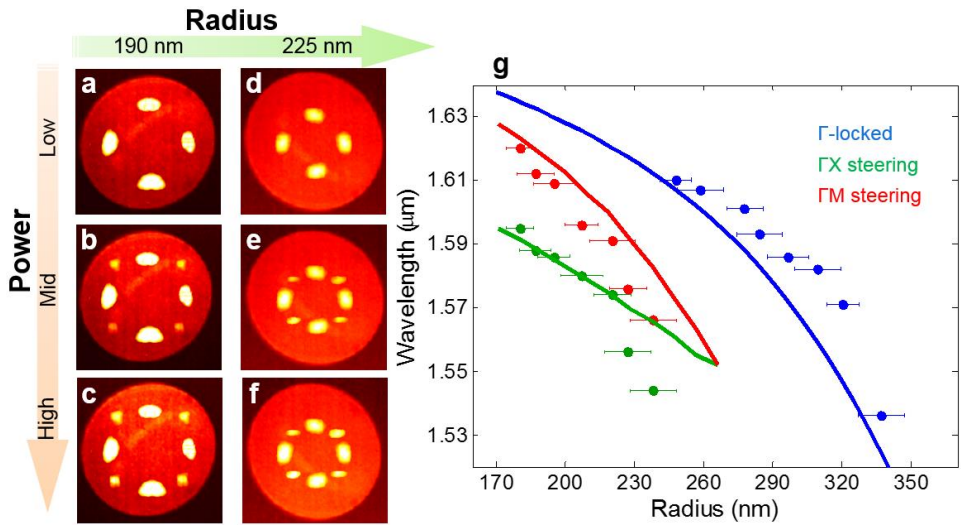

**Figure 4 | Topological charge conservation and scaling of the lasers.** (**a-f**) Reciprocal space images of the vortex laser when the only mode lasing is the one beaming along ΓX (mode 3) (a, d). This mode has a topological charge of +1, and, topology guarantees the existence of another mode with the opposite charge in the system, as charges are created in pairs. As the power is increased, it reaches the threshold of the mode beaming along ΓM (mode 2) that also starts to lase (b, e) as evident from the four additional bright spots (four-fold symmetry) in k-space appearing at 45° to the previous lasing spots. Increasing the pump power further, increases the emission power of the lasing modes, making them brighter (c, f). In our system, the threshold of the mode beaming along ΓM (mode 2, threshold$_{ΓM}$) is larger than the threshold of the mode beaming along ΓX (mode 3, threshold$_{ΓX}$) (see Supplementary Information). For pump powers smaller than threshold$_{ΓM}$ (threshold$_{ΓX}$ < threshold$_{ΓM}$), the mode beaming along ΓX is the only mode lasing in the device. The relative threshold of the modes is governed by their effective gain that depends not only on the modes distribution in the device but also the gain spectrum of the multiple quantum wells (see Supplementary Information). For $R$ = 190 nm the emission angle of mode 2 is 13.5° and the emission angle of mode 3 is 11.1°. For $R$ = 225 nm, the emission angle of modes 2 and 3 is 8°, demonstrating beaming, both along ΓX and ΓM. (**g**) Lasing wavelength as a function of the radius of holes from 180 nm to 350 nm. Each point corresponds to a device with a specific radius. Error bars indicate the standard deviation of radii measured from fabricated devices. Solid lines represent the wavelength of the singularities (resonances with large quality factor) of mode 1 (blue), mode 2 (red), and mode 3 (green) for different radii of the holes. The good agreement between theory and experiment constitutes an additional identification of the modes.